\documentclass[preprint2]{aastex6}

\usepackage{amsmath}

\shortauthors{Sheehan et al.}
\shorttitle{WL 17}

\begin{document}

\title{WL 17: A Young Embedded Transition Disk}
\author{Patrick D. Sheehan\altaffilmark{1}, Josh A. Eisner\altaffilmark{1}}
\affil{\altaffilmark{1}Steward Observatory, University of Arizona, 933 N. Cherry Avenue, Tucson, AZ, 85721}


\begin{abstract}
We present the highest spatial resolution ALMA observations to date of the Class I protostar WL 17 in the $\rho$ Ophiuchus L1688 molecular cloud complex, which show that it has a 12 AU hole in the center of its disk. We consider whether WL 17 is actually a Class II disk being extincted by foreground material, but find that such models do not provide a good fit to the broadband SED and also require such high extinction that it would presumably arise from dense material close to the source such as a remnant envelope. Self-consistent models of a disk embedded in a rotating collapsing envelope can nicely reproduce both the ALMA 3 mm observations and the broadband SED of WL 17. This suggests that WL 17 is a disk in the early stages of its formation, and yet even at this young age the inner disk has been depleted. Although there are multiple pathways for such a hole to be created in a disk, if this hole were produced by the formation of planets it could place constraints on the timescale for the growth of planetesimals in protoplanetary disks.
\end{abstract}

\section{Introduction}

Protoplanetary disks are the birthplaces of planets. Many protoplanetary disks have been found to have large central clearings. This was initially discovered by modeling disk SEDS \citep[e.g.][]{Strom1989,Espaillat2007}, but more recently these holes have been directly imaged with millimeter interferometers \citep[e.g.][]{Isella2010, Andrews2011}. These ``transition" disks have been hypothesized to be the result of planets carving holes in disks \citep[e.g.][]{DodsonRobinson2011}, although other physical processes such as photoevaporation and dust grain growth can also explain these holes \citep[e.g][]{Dullemond2005,Alexander2006}. Recently planets have been found hiding in the cavities, giving credibility to the idea that the holes are carved by planets \citep[e.g.][]{Sallum2015}. However, these transition disks have only been found in the older sample of Class II disks, which are thought to have ages of a few million years \citep[e.g.][]{Andre1994,Barsony1994}.

WL 17 is a M3 protostar in the L1688 region of the $\rho$ Ophiuchus molecular cloud \citep{Doppmann2005}, located a distance of 137 pc away \citep{OrtizLeon2017}. It has consistently been identified as a Class I protostar \citep{vanKempen2009,Enoch2009}, meaning it is younger than $\sim5\times10^5$ years and still embedded in envelope material from the collapsing molecular cloud \citep[e.g.][]{Evans2009}. The SED of WL 17 peaks in the mid-infrared, and shows a lack of optical emission that demonstrates that the source is highly extincted \citep{Enoch2009}. Low spatial resolution millimeter observations of WL 17 suggest the presence of large scale emission, likely from the remnants of a protostellar envelope \citep{vanKempen2009}.  Moreover, these same observations detected HCO$^+$ $J = 4-3$ emission towards WL 17 that is too bright to be associated with a disk. In addition, a survey of outflows in the L1688 region of Ophiuchus found that there is a weak outflow associated with WL 17 \citep{vanderMarel2013}. 
All of these signs point towards WL 17 being a young source that is still embedded in its natal envelope.

As such, it was observed as part of our ALMA survey of young embedded protostars in Ophiuchus \citep{Sheehan2017b}. However, upon imaging WL 17 we were surprised to find that it has a large hole in its center, suggesting a transition disk. While the highly reddened SED peaking in the mid-infrared clearly shows that WL 17 is embedded, it is not unprecedented to find disks that are extincted by the large scale cloud \citep[e.g.][]{Boogert2002,Brown2012}. Here we explore the nature of the medium extincting WL 17 to determine whether it is a young protostar still embedded in its natal envelope, which has cleared out a hole despite its young age, or whether it is an older, disk-only source that has been highly extincted by foreground dust.


\section{Observations \& Data Reduction}

\subsection{ALMA}

\floattable
\begin{deluxetable}{lccc}
\tablenum{1}
\tablecaption{Log of ALMA Observations}
\label{table:alma_obs}
\tablehead{\colhead{Observation Date} & \colhead{Baselines} & \colhead{Total Integration Time} & \colhead{Calibrators} \\
\colhead{(UT)} & \colhead{(m)} & \colhead{(s)} & \colhead{(Flux, Bandpass, Gain)}}
\startdata
Oct. 31 2015 & 84$\,$-$\,$16,200 & 169 & 1517-2422, 1625-2527 \\
Nov. 26 2015 & 68$\,$-$\,$14,300 & 169 & 1517-2422, 1625-2527 \\
Apr. 17 2016 & 15$\,$-$\,$600 & 58 & 1733-1304, 1427-4206, 1625-2527 \\
\enddata
\end{deluxetable}

WL 17 was observed with ALMA in three tracks from 31 October 2015 to 17 April 2016, with baselines ranging from 14 m -- 15.3 km. The observations were done with the Band 3 receivers, and the four basebands were tuned for continuum observations centered at 90.5, 92.5, 102.5, 104.5 GHz, each with 128 15.625 MHz channels for 2 GHz of continuum bandwidth per baseband. In all the observations had 8 GHz of total continuum bandwidth. We list details of the observations in Table \ref{table:alma_obs}.

\begin{figure}[b!]
\centering
\figurenum{1}
\includegraphics[width=3in]{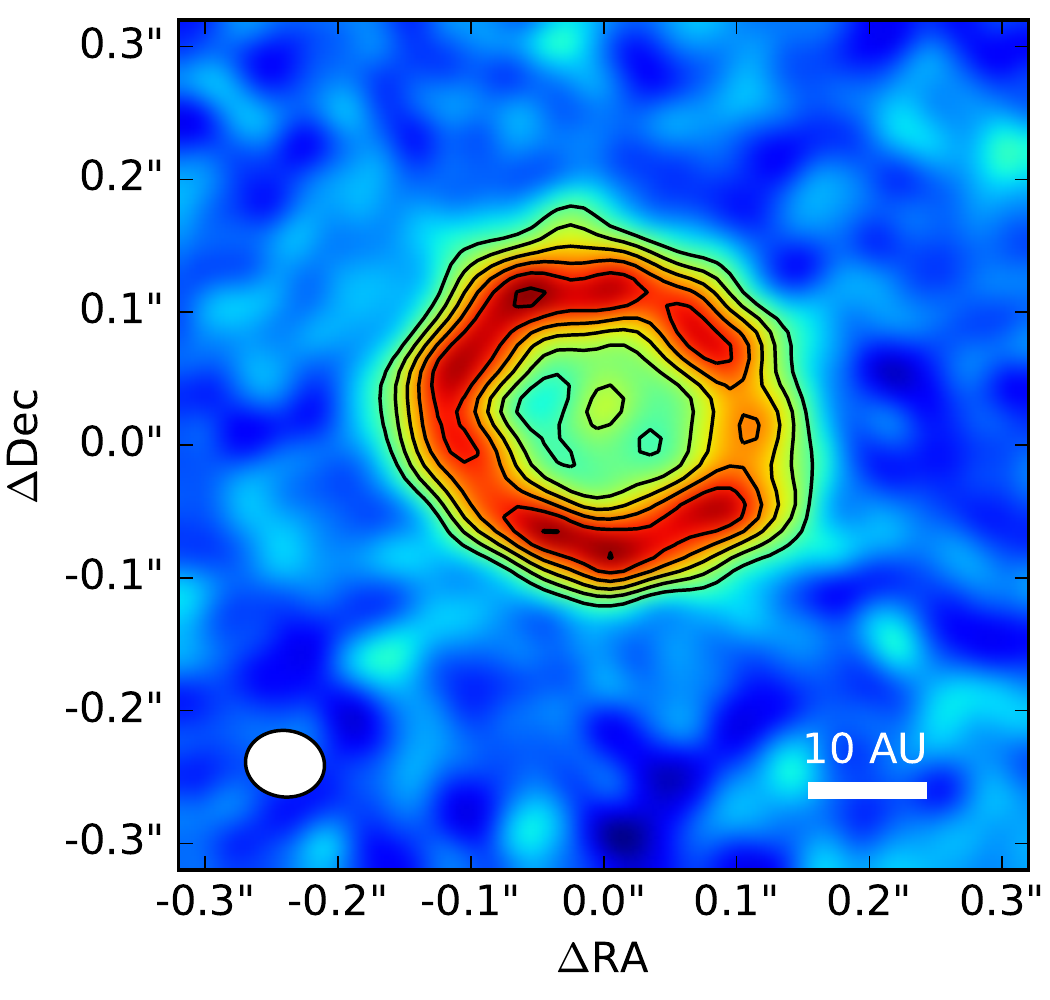}
\caption{ALMA 3 mm map of WL 17 showing a clear ring-like structure. The synthesized beam size is 0.06" by 0.05" with a P.A. of 81.9$^{\circ}$. Contours begin at 4$\sigma$ and subsequent contours are every additional 2$\sigma$, with $1\sigma = 36$ $\mu$Jy. The emission interior to the ring does not drop to zero, but rather falls to a $4\sigma$ level at the inner edge of the ring. At the center of the ring the emission rises to a $6\sigma$ level. This may indicate the presence of material remaining in the cleared out region.}
\label{fig:alma_data}
\end{figure}

We reduce the data in the standard way with the \texttt{CASA} software package and the calibrators listed in Table \ref{table:alma_obs}. After calibrating, we image the data by Fourier transforming the visibilities with the \texttt{CLEAN} routine. We use Briggs weighting with a robust parameter of 0.5, which provides a good balance between sensitivity and resolution, to weight the visibilities. The resulting image has a beam of size 0.06" by 0.05" with a P.A. of 81.9$^{\circ}$. We show the resulting image in Figure \ref{fig:alma_data}, and the azimuthally averaged visibility amplitudes in Figure \ref{fig:models}. The rms of the image is 36 $\mu$Jy/beam. All analysis is done directly to the un-averaged two dimensional visibilities.

\subsection{SED from the Literature}

\begin{figure*}[t]
\centering
\figurenum{2}
\includegraphics[width=6.5in]{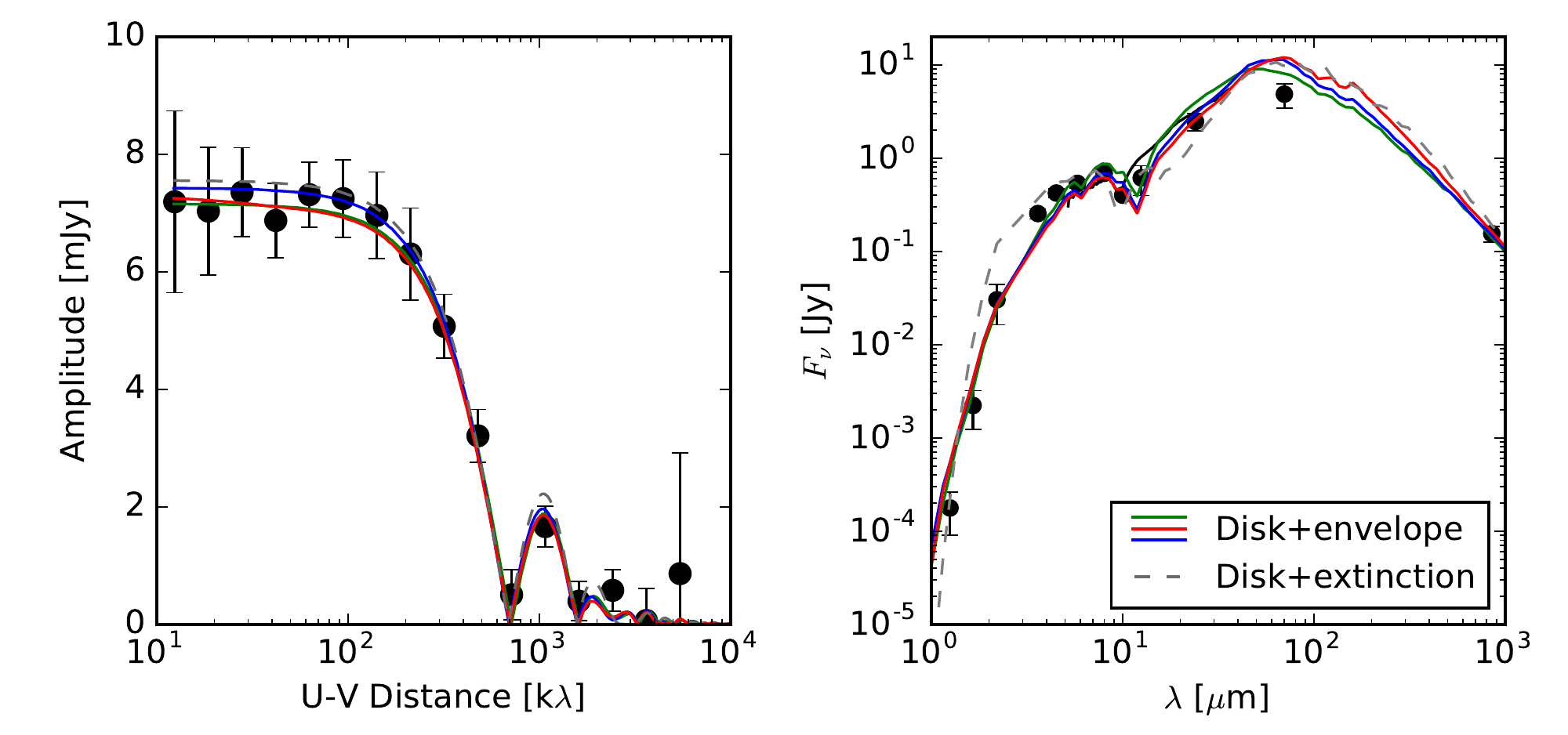}
\caption{Examples of models that fit the combined WL 17 3 mm visibilities ({\it left}) + SED ({\it right}) dataset. We show our broadband SED and the 1D azimuthally averaged visibilities as black points, and the IRS spectrum is shown as a black line. In gray we show the disk+foreground extinction model that does not fit the data well. In red, green, and blue red and blue we show three possible disk+envelope models that can well fit the data with a range of values for the envelope mass and radius. Parameter values for these models, as well as metrics to assess the quality of the fits, are listed in Table \ref{table:model_params}.}
\label{fig:models}
\end{figure*}

We compile a broadband spectral energy distribution (SED) for WL 17 from a thorough literature search. We show the SED in Figure \ref{fig:models}. The data includes {\it Spitzer} IRAC and MIPS photometry as well as fluxes from the literature at a range of wavelengths \citep{Wilking1983,Lada1984,Greene1992,Andre1994,Strom1995,Barsony1997,Johnstone2000,Allen2002,Natta2006,Stanke2006,AlvesdeOliveira2008,Jorgensen2008,Padgett2008,Wilking2008,Evans2009,Gutermuth2009,Barsony2012}. We exclude WISE photometry because the fluxes are inconsistent with the IRAC and MIPS fluxes. This is because the WISE beam is larger than the Spitzer beam, and may cause confusion with nearby sources. The IRAC and MIPS flux measurements were also independently reproduced by two different groups using separate datasets \citep{Evans2009,Gutermuth2009}, so we believe these measurements to be reliable.

We also include the the SL, SH, and LH calibrated Spitzer IRS spectrum from the CASSIS database in our SED\citep{Lebouteiller2011,Lebouteiller2015}. We find that we need to scale the IRS spectrum by a factor of 3 to align it with the IRAC/MIPS photometry for the system. When scaled up the IRS spectrum also nicely matches ground-based 10 $\mu$m photometry of the silicate absorption feature. This factor may be needed due to issues in the flux calibration or the pointing towards the source.

For the purposes of assessing the quality of model fits to the SED we assume a 10\% uncertainty on all flux measurements when computing $\chi^2$. We also sample the IRS spectrum at 25 points evenly spaced across the spectrum to minimize the number of individual wavelengths at which radiative transfer flux calculations, which can be time intensive, must be done.

\section{Results}

Our 3 mm map of WL 17, shown in Figure \ref{fig:alma_data}, shows a well detected, compact source with a hole measuring $\sim0.2''$ in diameter in the center. At the distance of Ophiuchus, which we assume to be 137 pc, the hole is 27 AU across ($\sim$13 AU in radius). Emission at the center of the hole peaks at $\sim~250$ $\mu$Jy, which suggests that there may still be material remaining inside the transition disk cavity. Alternatively this could be emission from magnetic activity at the surface of a young star.

Studies that have found holes in the centers of many other protoplanetary disks, dubbed ``transition disks" \citep{Espaillat2007,Isella2010,Andrews2011}. Transition disks are typically found in the population of Class II protoplanetary disks, which represents older disks that are no longer embedded in envelopes. Unlike these previous detections, WL 17 has an SED (shown in Figure \ref{fig:models}) that peaks at mid-infrared wavelengths and looks very much like a Class I source. WL 17 must be embedded in some obscuring material, but stars form in giant clouds of gas and dust, so it is reasonable to think that WL 17 could be a Class II source made to look like a Class I by foreground extinction from this cloud. Transition disks have been previously found with significant amounts of extinction from foreground material \citep[e.g.][]{Boogert2002,Brown2012}. It is also possible to mistake edge on Class II disks as Class I sources \citep[e.g.][]{Chiang1999}.

\begin{figure*}[t]
\centering
\figurenum{3}
\includegraphics[width=6in]{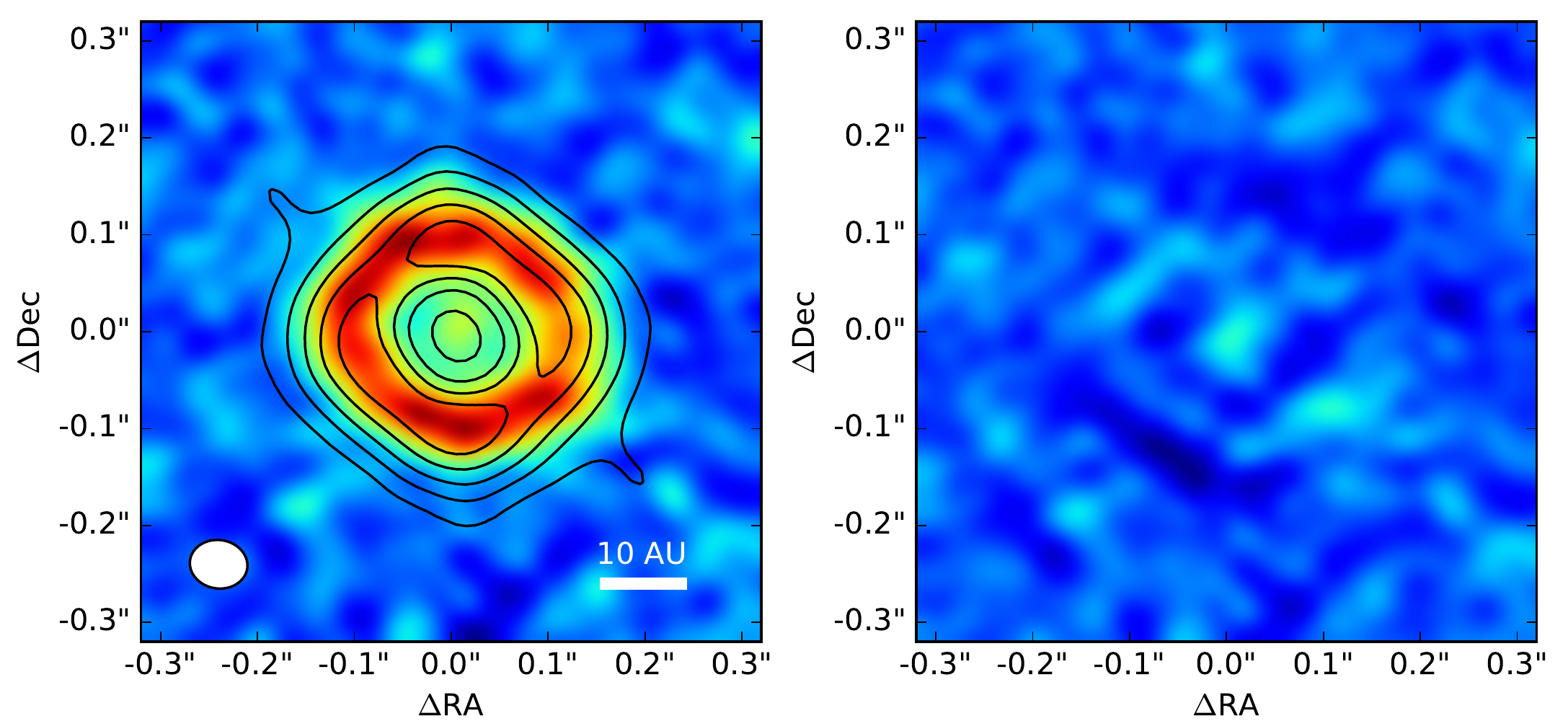}
\caption{({\it left}) ALMA 3 mm map of WL 17 with the best fit disk+envelope model as contours to demonstrate the good match of the model to the data in the image plane. ({\it right}) Residual map produced by subtracting our best fit model from the 3 mm map in the visibility plane and inverting to produce an image. The peak residual is at a 5$\sigma$ level, but the rest are $<3\sigma$. The large residual level comes from the somewhat clumpy structure seen in the image. We employ a fairly simple model that assumes the disk structure is smooth, so we cannot expect to fully reproduce this clumpy structure with our model.}
\label{fig:alma_model}
\end{figure*}

As such, a disk model that includes foreground extinction is a good first guess for attempting to reproduce the combined ALMA 3 mm visibilities and broadband SED dataset. To do so we use detailed radiative transfer models, run using the Monte Carlo radiative transfer codes \texttt{RADMC-3D} and \texttt{Hyperion} \citep{Robitaille2011,Dullemond2012}, to produce synthetic visibilities and SEDs and attempt to match the data with the models. We give a brief description of the models here, but refer to \citet{Sheehan2014} for a more detailed account.

Our model assumes a central protostar with a M3 spectral type \citep[$T = 3400 K$;][]{Doppmann2005}, although we allow the luminosity of the protostar, $L_*$, to vary. We include a disk with a power law surface density, 
\begin{equation}
\rho = \rho_0 \left(\frac{R}{R_0}\right)^{-\alpha} \, \exp\left(-\frac{1}{2}\left[\frac{z}{h(R)}\right]^2\right)
\end{equation}
where $R$ and $z$ are in cylindrical coordinates. $h(R)$ is the disk scale height at a given radius,
\begin{equation}
h(R) = h_0 \left(\frac{R}{1 \text{ AU}}\right)^{\beta}.
\end{equation}
We truncate the disk at some outer radius, $R_{disk}$, and specify a gap radius, $R_{gap}$, inside of which the density is decreased by a multiplicative factor, $\delta$. $\alpha$, $\beta$, $h_0$, and the disk mass $M_{disk}$ are also left as free parameters, as are the inclination and position angle of the system. We supply the model with dust opacities from \citet{Sheehan2014}, but allow the maximum size of the dust grain size distribution, $a_{max}$, to vary. We extinct the synthetic SED by some number of K-band magnitudes, $A_K$, using the \citet{McClure2009} extinction law. The model visibilities are unaffected by this extinction because extinction at millimeter wavelengths from foreground dust is negligible. Moreover, our millimeter observations resolve out large scale emission from the foreground cloud.

We show the best fit disk+extinction model in Figure \ref{fig:models}, and list the best fit parameter values in Table \ref{table:model_params}. Although the model well reproduces the 3 mm visibility profile, it cannot produce a good fit to the SED as it under-predicts the mid-infrared flux. This is because, with an inner radius of 12 AU, there is not enough hot material close to the star to overcome foreground extinction and produce the necessary mid-infrared flux. The model also slightly over-predicts the amount of near-infrared flux. Moreover, $A_K \sim 4$ ($A_V \sim 30$ for the \citet{McClure2009} extinction law) is needed to properly extinct the near-infrared SED. \citet{Boogert2002} found two foreground clouds that contribute $A_V\sim11$ in the region near WL 17, but this is not enough to explain the $A_V\sim30$ needed to match the SED. Such high extinction seems unlikely to come from foreground extinction from nearby star forming regions.

A more natural explanation for the extinction towards WL 17 is that it is still a young disk embedded in its natal envelope. To test this hypothesis we consider a disk+envelope density distribution model to see whether it can reproduce our dataset. We use the same prescription for the disk, but embed the disk in a rotating collapsing envelope \citep{Ulrich1976}. The density profile for the envelope is given by,
\begin{equation}
\footnotesize
\rho = \frac{\dot{M}}{4\pi}\left(G M_* r^3\right)^{-\frac{1}{2}} \left(1+\frac{\mu}{\mu_0} \right)^{-\frac{1}{2}} \left(\frac{\mu}{\mu_0}+2\mu_0^2\frac{R_c}{r}\right)^{-1},
\end{equation}
where $\mu = \cos \theta$ and $r$ and $\theta$ are in spherical coordinates. Here the mass and radius of the envelope ($M_{env}$ and $R_{env}$) are left as free parameters and the envelope is truncated at an inner radius of 0.1 AU. The critical radius $R_c$ represents the radius inside of which the density distribution flattens into a disk-like structure, with the majority of material being deposited at $R_c$. This makes the most sense physically if $R_c \sim R_{disk}$, so we provide this constraint to our modeling. We still allow for a small amount of extinction towards WL 17 in the disk+envelope model because of the known foreground clouds in the region.

Our disk+envelope model is able to produce good fits to the combined 3 mm visibilities and broadband SED dataset. We show a few examples of these fits in Figure \ref{fig:models}. These models were found by taking the disk+extinction model disk parameters, adding an envelope, and adjusting the parameters by hand to find models that produce better $\chi^2$ values. These models are not ``best fits" because no optimization was done, but their $\chi^2$ values (see Table \ref{table:model_params}) are clearly better than that of the disk+extinction model. 

In Figure \ref{fig:alma_model} we compare the 3 mm ALMA map with a representative image of the disk+envelope model, which we produced by sampling a synthetic 3 mm image from our radiative transfer model at the same spatial frequencies as the ALMA data before making the image. Unlike the disk+extinction model, which under-predicts the mid-infrared flux, the disk+envelope model is better able to fit the mid-infrared spectrum of WL 17. This is because the envelope allows for more hot material close in to the protostar, boosting the mid-infrared flux. 

\floattable
\begin{deluxetable}{lcccccccccccccccc}
\tabletypesize{\scriptsize}
\tablenum{2}
\tablecaption{Model Parameters}
\label{table:model_params}
\tablehead{\colhead{Model} & \colhead{$L_{*}$} & \colhead{$M_{disk}$} & \colhead{$R_{in}$} & \colhead{$R_{disk}$} & \colhead{$h_0$\tablenotemark{a}} & \colhead{$\gamma$\tablenotemark{b}} & \colhead{$\beta$\tablenotemark{c}} & \colhead{$\delta$\tablenotemark{d}} & \colhead{$M_{env}$} & \colhead{$R_{env}$} & \colhead{$i$} & \colhead{$P.A.$} & \colhead{$a_{max}$} & \colhead{$A_K$} & \colhead{$\chi_{vis}^2$} & \colhead{$\chi_{SED}^2$} \\
\colhead{ } & \colhead{$[L_{\odot}]$} & \colhead{$[M_{\odot}]$} & \colhead{[AU]} & \colhead{[AU]} & \colhead{[AU]} & \colhead{} & \colhead{} & \colhead{} & \colhead{$[M_{\odot}]$} & \colhead{[AU]} & \colhead{$[\,^{\circ}\,]$} & \colhead{$[\,^{\circ}\,]$} & \colhead{[mm]} & \colhead{[mag]} & \colhead{} & \colhead{}}
\startdata
Disk+envelope (green) & 0.5 & 0.05 & 11.6 & 22.7 & 0.15 & 0.0 & 0.75 & 0.011 & $3\times10^{-5}$ & 25 & 28 & 82.4 & 10 & 0.5 & 1067 & 112 \\
Disk+envelope (blue) & 0.5 & 0.04 & 11.6 & 22.7 & 0.15 & 0.0 & 0.75 & 0.011 & $3\times10^{-4}$ & 100 & 28 & 82.4 & 10 & 0.5 & 1070 & 145 \\
Disk+envelope (red) & 0.5 & 0.035 & 11.6 & 22.7 & 0.2 & 0.0 & 0.75 & 0.01 & 0.003 & 600 & 28 & 82.4 & 10 & 0.75 & 1071 & 172 \\
Disk+extinction & 6.2 & 0.06 & 11.6 & 22.7 & 0.11 & -0.24 & 1.0 & 0.02 & \nodata & \nodata & 28 & 82.4 & 0.3 & 4.2 & 1078 & 325 \\
\enddata
\tablenotetext{a}{$h_0$ is the disk scale height at 1 AU.}
\tablenotetext{b}{$\gamma$ is the disk surface density power law exponent.}
\tablenotetext{c}{$\beta$ is the power law exponent which describes how the disk scale height varies with radius (see Equation 2).}
\tablenotetext{d}{$\delta$ is the factor by which the disk density is reduced inside of the gap.}
\end{deluxetable}

There is a significant degeneracy between envelope mass and radius in these models; both large, high mass and small compact envelopes can produce the extinction needed to match the SED. Our millimeter observations resolve out scales larger than $\sim$20'', or radii larger than $\sim$1300 AU, so our data is not sensitive to large scale envelope structure. Moreover, the visibilities lack the sensitivity at intermediate scales to detect faint emission from a more compact envelope. As such, our modeling cannot well distinguish between compact low mass envelopes and larger and more massive envelopes.

\section{Discussion \& Conclusion}

In order to provide a quantitative assessment of the quality of fit of our models, we have computed the $\chi^2$ value for each of the models listed in Table \ref{table:model_params} for both the SED and the 3 mm visibilities. We find that for all four models, including both our disk+extinction and disk+envelope models, the quality of the fit to the 3 mm visibilities is indistinguishable; all models are able to reproduce the observed 3 mm visibilities of WL 17. The disk+extinction model, however, has a much worse $\chi^2$ value for the SED than the disk+envelope models. Only the disk+envelope models can well reproduce both the visibilities and the broadband SED simultaneously. The disk+extinction model cannot simultaneously reproduce both datasets, and moreover the best fit disk+extinction model requires $A_V\sim30$, which is quite high for foreground extinction.

The good fit of the disk+envelope models, as well as the high extinction required of the disk+extinction model, indicate that the extinction seen towards WL 17 is the result of it being embedded in an envelope of dusty material. This matches nicely with previous studies of the system, discussed above, that have hinted at its youth \citep{vanKempen2009,vanderMarel2013}.

As such, we suggest that WL 17 is a young source still embedded in the remnants of its natal envelope. It may be, if the envelope remnants are low-mass, that the system is in the process of shedding the final layers of envelope and will soon be exposed as a more traditional transition disk system. However, the presence of even a low-mass remnant envelope indicates youth.  Moreover, substantially more massive envelopes cannot be ruled out.

Regardless of the exact nature of the envelope, the discovery of a transition disk still embedded in its envelope raises interesting questions. There are a few explanations for such a hole, including photoevaporation of the inner disk by the central protostar, dust grain growth in the inner disk, and a dynamical clearing of the inner disk by large bodies \citep[e.g][]{Dullemond2005,Alexander2006,DodsonRobinson2011}. Embedded protoplanetary disks are thought to be only a few hundred thousand years old \citep[e.g.][]{Andre1994,Barsony1994}, so any explanation of the presence of the hole must be compatible with a young age.

Photoevaporation tends to be ineffective early in the lifetimes of disks, when the accretion rate exceeds the photoevaporation rate; furthermore, once a gap is opened, the disk is dispersed quickly \citep[e.g.][]{Alexander2006}. Photoevaporation models that include the influence of FUV and X-ray photons produce significantly higher photoevaporation rates, and could explain the presence of a large hole early in the lifetime of a disk \citep[e.g.][]{Gorti2009a,Owen2010,Armitage2011}. Still, these models require low accretion rates to be effective, and if this system is embedded in an envelope, the accretion rate is unlikely to be low.

Dust grain growth in the inner disk could be possible. Our millimeter observations show a dearth of millimeter sized bodies within the hole, but it is possible that this hole is indicating that even larger planetesimals have formed here. That said, dust grain growth may have challenges reproducing the sharp inner edge seen in Figure \ref{fig:alma_data} \citep[e.g.][]{Birnstiel2012}.

If the disk is dynamically cleared, it may be that WL 17 is a compact binary system. Radial velocity searches of the system have been done and no evidence of a companion has been found \citep{VianaAlmeida2012}, although the limits are not strong. For a companion just inside the disk at 10 AU and a sensitivity to changes in radial velocity of $\sim4-6$ km s$^{-1}$, we estimate an upper limit on the mass of a companion of $<0.25-0.4$ M$_{\odot}$, although the true limit is likely worse given the sparse sampling of the data. Another, perhaps more exciting possibility, is that this hole may be cleared out by a planet or multiple planets.

It may seem surprising to find a young disk with a large hole as it may require the presence of planets at a very young age. Planets can, however, form quickly in massive disks \citep[e.g.][]{Pollack1996}. Indeed, multiple gaps have been found in the disk of HL Tau, another young and possibly embedded protostar likely between the Class I and II stages \citep{Brogan2015}. The gaps in the HL Tau disk, however, can be produced by Saturn-mass objects \citep[e.g.][]{Dong2015} whereas transition disk holes like the one seen in WL 17 may need planets of a Jupiter-mass or larger \citep[e.g.][]{DodsonRobinson2011}. The existence of gaps and holes in young embedded disks seems to indicate that the processes that govern planet formation must happen quite quickly, as planets must grow to large enough masses to clear out holes in their disks in a short amount of time.

\acknowledgements

We would like to thank Nienke van der Marel and George Rieke for helpful conversations and feedback. This work was supported by NSF AAG grant 1311910. This paper makes use of the following ALMA data: ADS/JAO.ALMA\#2015.1.00761.S. ALMA is a partnership of ESO (representing its member states), NSF (USA) and NINS (Japan), together with NRC (Canada), NSC and ASIAA (Taiwan), and KASI (Republic of Korea), in cooperation with the Republic of Chile. The Joint ALMA Observatory is operated by ESO, AUI/NRAO and NAOJ The National Radio Astronomy Observatory is a facility of the National Science Foundation operated under cooperative agreement by Associated Universities, Inc.

\end{document}